\documentclass[a4paper,11pt,abstracton]{scrartcl} 
\usepackage{amsmath}
\usepackage{graphicx}

\begin{document}

\title{Optical Properties and Modal Gain of InGaN Quantum Dot Stacks}
\date{\begin{footnotesize}accepted 2008/11/27 for publication in physica status solidi (c)      \end{footnotesize}}
\author{Joachim Kalden, Kathrin Sebald, and J\"urgen Gutowski\\Christian Tessarek, Timo Aschenbrenner, Stephan Figge, and Detlef Hommel\\[1ex]Institute of Solid State Physics, University of Bremen\\P.O. Box 330 440, 28334 Bremen, Germany}
\maketitle
\begin{abstract}
We present investigations of the optical properties of stacked InGaN quantum dot layers and demonstrate their advantage over single quantum dot layer structures. Measurements were performed on structures containing a single layer with quantum dots or threefold stacked quantum dot layers, respectively. A superlinear increase of the quantum dot related photoluminescence is detected with increasing number of quantum dot layers while other relevant GaN related spectral features are much less intensive when compared to the photoluminescence of a single quantum dot layer. The quantum dot character of the active material is verified by microphotoluminescence experiments at different temperatures. For the possible integration within optical devices in the future the threshold power density was investigated as well as the modal gain by using the variable stripe length method. As the threshold is 670\,kW/cm\textsuperscript{2} at 13\,K, the modal gain maximum is at 50\,cm\textsuperscript{-1}. In contrast to these limited total values, the modal gain per quantum dot is as high as 10\textsuperscript{-9}cm\textsuperscript{-1}, being comparable to the II-VI and III-As compounds. These results are a promising first step towards bright low threshold InGaN quantum dot based light emitting devices in the near future.
\end{abstract}
\section{Introduction}
Semiconductor quantum dots (QDs) have attracted much interest in recent years due to the high stability of their emission properties when changing external conditions~\cite{senes_highT_2007,sebald_optical_2008}. Due to the three-dimensional electron confinement, QDs as active material allow for the realization of low-threshold light emitting and laser diode structures~\cite{arakawa_thresT_1982,Asada_QDlaser_1986,kruse_integration_2008}. Furthermore, it is possible to employ the quantum-confined Stark effect (QCSE) occurring due to the built-in piezo-electric field~\cite{chow_microscopic_1999,bretagnon:113304} to shift the PL to longer wavelengths, since the transparency carrier density is reduced compared to quantum well (QW) based devices~\cite{chow_theory_2002}. However, it is necessary to increase the QD density by using stacked layers to achieve sufficient active material for appropriate lasing operation. This method has been very successful for the group-III arsenides~\cite{As-lasing,schmidt_prevention_1996} and the II-VI wide bandgap semiconductors~\cite{sebald_optical_2002}. In addition, stacking of the QD layers  is expected to result in an increasing uniformity of the QD size throughout the layering process~\cite{mateeva:3233}, resulting in an increasing modal gain~\cite{qiu:3570}. 
\section{Sample structure and experimental setup}
The investigated samples were grown by metal-organic vapor pha\-se epitaxy. On the (0001)-sapphire substrate a 2\,$\mu$m thick GaN buffer layer was grown, followed by a 540\,nm Al\textsubscript{0.12}Ga\textsubscript{0.88}N cladding layer. Subsequently, the lower wave\-guide layer was realized by 104\,nm GaN, followed by the threefold stack of InGaN QD layers. Every QD layer was grown by a two-step method~\cite{yamaguchi_twostepgrowth_2006}, applying a 1.5\,nm nucleation layer consisting of In\textsubscript{0.18}Ga\textsubscript{0.82}N overgrown by a 5\,nm formation layer of In\textsubscript{0.08}Ga\textsubscript{0.92}N, and finally capped by 14\,nm GaN. The GaN capping of the top QD layer is 117\,nm thick, representing the upper waveguide layer as well. The upper cladding is omitted since the index contrast from GaN (n $\approx$ 2.5) to air (n $\approx$ 1) is sufficiently large.\\[1ex]
Reference samples without cladding layers are investigated via microphotoluminescence ($\mu$PL) experiments with a spatial resolution of about 2\,$\mu$m given by the laser spot diameter. To further reduce the active area, mesa structures are processed by focused ion beam (FIB) milling, which then can be individually studied via $\mu$PL. The excitation source is a HeCd laser operating in cw mode at 325\,nm (3.815\,eV). A reflecting microscope objective with a numerical aperture of 0.5 is used to focus the laser beam on the sample and to detect its PL signal. The threshold of the waveguide structures is determined by power density dependent measurements while the modal gain was deduced via the variable stripe length (VLS) method~\cite{shaklee_VLS_1973,roewe_influence_2003}. For these experiments, the excitation source is a XeCl excimer laser operating at 308\,nm (4.025\,eV) in pulsed mode with a pulse width of 5\,ns  pumping a dye laser operating at 360\,nm (3.444\,eV). To achieve a well defined stripe-like spot as demanded for VLS experiments, the beam is shaped via an rectangular pin hole and a cylindrical lens. The variable stripe length is conrolled by stepper-driven razor blades in the range from 0 -- 500\,$\mu$m.
\section{Results and discussion}
\begin{figure}[tb]
 \includegraphics[width=0.98\textwidth]{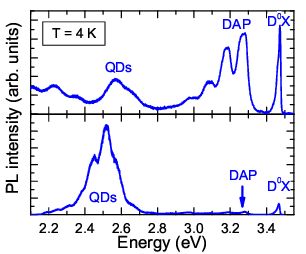}
 \caption{Comparison of the PL of a single QD layer (top) and a threefold QD stack (bottom). The QD PL is significantly increased for the stack, while D\textsuperscript{0}X and DAP intensities are less intensive than for the single layer (see text).}
 \label{fig:vglPL}
\end{figure}

\begin{figure}[tb]
 \includegraphics[width=0.98\textwidth]{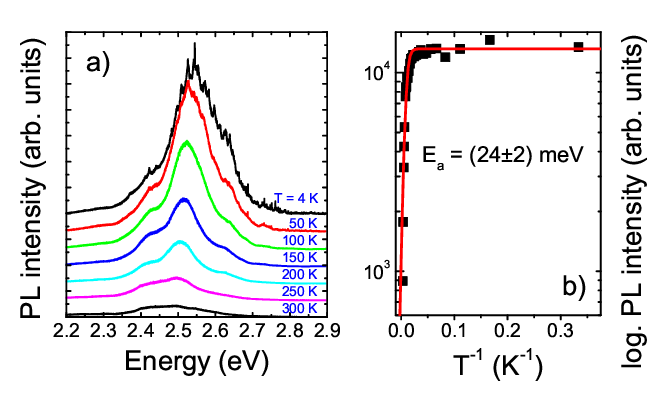}
 \caption{a) PL of an active area of about 1.2\,$\mu$m\textsuperscript{2} for different temperatures. b) Arrhenius plot of QD band maximum~\cite{bacher_influence_1991}. E\textsubscript{a} is the activation energy.}
 \label{fig:activation}
\end{figure}

\begin{figure}[tb]
 \includegraphics[width=0.98\textwidth]{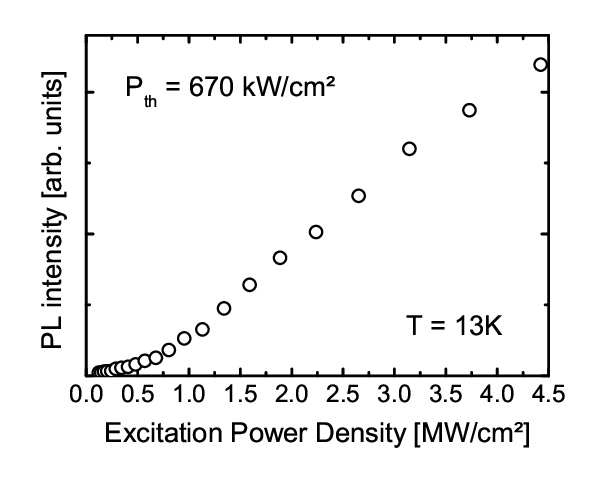}
 \caption{Power density dependent PL intensity at 13\,K. The threshold power density P\textsubscript{th} is found to be 670\,kW/cm\textsuperscript{2}.}
 \label{fig:thres}
\end{figure}

Photoluminescence measurements are presented in Fig.~\ref{fig:vglPL} for a single QD layer sample as well as a threefold QD stack. Both spectra show a modulation of the QD emission band which is caused partly by Fabry-Perot oscillations between the interfaces air/GaN and GaN/sapphire. Additionally, the spectral position of each layer in the stacked sample may slightly differ which could also yield some local maxima in the QD emission band. For the stacked sample the QD emission intensity dominates the PL spectrum relative to the donor-bound exciton (D\textsuperscript{0}X) and the donor-acceptor-pair (DAP) recombination band intensity. In order to quantitatively describe the relative intensity of the QD PL, the integrated PL intensities of D\textsuperscript{0}X and QD emission for both samples are compared. As we find the QD emission to be 3 times more intensive than the D\textsuperscript{0}X for the single layer, it is determined to be 65 times stronger for the QD stack. This large increase cannot be explained just by the superposition of three independent QD layers, thus it may be due to a decreased exciton lifetime reported in vertically coupled QD stacks~\cite{ledentsov_direct_1996} leading to a relatively higher PL intensity. For QD stacks, it is further reported that the QD size slowly increases layer by layer, leading to an overall redshift of the QD PL~\cite{hospodkova_photoluminescence_2007} as seen in Fig.~\ref{fig:vglPL}. However, up to now no transmission electron microscopy studies have been performed to prove this assumption of vertical coupling.\\[0.95ex]
When the active area of the stack sample is reduced to about 1.2\,$\mu$m\textsuperscript{2} via FIB, $\mu$PL reveals that the QD ensemble emission band splits up, and several sharp lines become visible at 4\,K, indicating the QD origin of the PL emission. Temperature dependent experiments show that the emission can still be traced up to room temperature for this small amount of active material (Fig.~\ref{fig:activation}a), demonstrating the high PL stability with respect to temperature increase of the stacked QD layers. This is confirmed by an activation energy of about 24\,meV deduced by using a thermal activation model~\cite{bacher_influence_1991} as shown in Fig.~\ref{fig:activation}b.\\[0.95ex]
Raising the excitation power density, the PL intensity of the stack sample increases decently up to a critical excitation power density usually called threshold density. A further raise of the excitation po\-wer den\-si\-ty leads to a steeper increase (Fig.~\ref{fig:thres}). The PL emission also gets spectrally narrower and stimulated emission sets in (not shown here). From measurements with different excitation power densities, this threshold density is deduced to 670\,kW/cm\textsuperscript{2} at 13\,K. For higher temperatures, this value is expected to be increased due to nonradiative recombination processes such as elec\-tron-phonon scattering which raise the losses that have to be overcome before optical gain is achieved. Also the crystalline quality of the GaN spacer layers grown at low temperatures in order to not dissolve the QDs could raise the threshold. Although the threshold power density for QD based devices is predicted~\cite{Asada_QDlaser_1986} and for the group-III arsenides demonstrated~\cite{liu_extremely_1999} to be much smal\-ler than for QW devices, it is here found to be still higher compared to InGaN QW structures~\cite{swietlik_comparison_2007}. This may be due to the fact that the QD layers contain much less active material than QW structures, since the QD density is limited to 5$\times$10\textsuperscript{9}cm\textsuperscript{-2} per layer. For CdSe QD stacks containing five layers with a QD density of up to 10\textsuperscript{11}cm\textsuperscript{-2} per layer, stimulated emission was demonstrated for 32 to 60 kW/cm\textsuperscript{-2}~\cite{sebald_optical_2002} while threshold densities of below 1kW/cm\textsuperscript{-2} were recently reported for arsenide based structures at room temperature~\cite{germann_temperature-stable_2008}.\\[0.95ex]
The modal gain of the waveguide structure is investigated via the VLS method. A rectangular laser spot with a fixed width and a variable length is focussed on the sample. The PL is detected for several stripe lengths. From these data the modal gain can be deduced~\cite{shaklee_VLS_1973,roewe_influence_2003}. For guaranteeing success of the VLS method, saturation effects have to be avoided. One reason for saturation effects can be that the stripe length becomes too large~\cite{vehse_optical_2001,kyhm:3434}. In this case the light generated on the stripe end opposite to the emitting edge will no longer be amplified since all excited states have already been depleted by photons coming from nearer distances. \\[0.95ex] 
Here, the modal gain maximum is g\textsubscript{mod}~=~50\,cm\textsuperscript{-1} at 12\,K, and the saturation stripe length is measured to L\textsubscript{s}~=~150\,$\mu$m. These values are mainly limited by the background absorption visible at lower energies where no absorption should occur in the ideal case. The product of modal gain maximum and saturation length is g\textsubscript{mod}L\textsubscript{s}~=~0.75, which is rough\-ly one order of magnitude less compared to QW structures, where the product has been measured to be 4 -- 10~\cite{vehse_optical_2001,mickevicius_saturated_2006}. InGaN QWs reach values of g\textsubscript{mod}~=~180 -- 300\,cm\textsuperscript{-1} and L\textsubscript{s}~=~250 -- 350\,$\mu$m~\cite{swietlik_comparison_2007}.
However, as the QD density is taken into account by calculating the modal gain per QD, it is found that our results are comparable with established materials~\cite{sebald_optical_2002}, yielding 10\textsuperscript{-9}cm\textsuperscript{-1} per QD. This gives evidence to the fact that rather the surrounding structure needs further improvement as well as the filling factor being still very low. For structures with higher QD densities up to 9$\times$10\textsuperscript{10}cm\textsuperscript{-2} per layer, a significantly lower threshold has recently been presented~\cite{wang_growth_2008}. Hence it should be possible to improve our results and achieve an even higher modal gain by increasing the QD density in our samples.
\begin{figure}[tb]
 \includegraphics[width=0.98\textwidth]{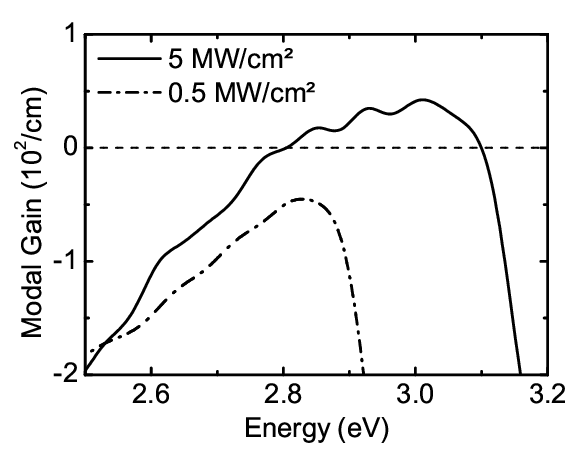}
 \caption{Modal gain spectrum at 12\,K calculated from VLS measurements with stripe lengths up to 150\,$\mu$m.}
 \label{fig:gain}
\end{figure}
\section{Conclusions}
Gallium nitride based waveguide structures con\-tai\-ning stacked quantum dot layers as active region grown by metal-organic vapor-phase epitaxy were presented and analyzed in detail. Photoluminescence experiments documented the quantum dot character of the stacked layers' spectral features. Temperature dependent measurements revealed a slight variation of the photoluminescence up to room temperature, which is a crucial fact for commercial applications in the future. The threshold density of the structure was found to be 670\,kW/cm\textsuperscript{2}. Modal gain of up to 50\,cm\textsuperscript{-1} was detected at low temperatures while a saturation length of 150\,$\mu$m was identified. Nevertheless, the modal gain per quantum dot of 10\textsuperscript{-9}cm\textsuperscript{-1} is found to be in the same regime as reported for the well established II-VI and III-As compounds. The findings presented are promising for future applications of light emitting devices based on InGaN quantum dot stacks. Next challenges are a further increase of the filling factor by stacking more layers and increasing the density per layer as well as further improvement on the structural quality to lower the background absorption in the structure. Another subsequent step in the future is the embedding of nitride based QD stacks in vertical emitting microcavity structures to combine the advantages of quantum dot stacks with features like a symmetric beam profile as well as single mode operation, the latter leading to a further threshold reduction~\cite{shore_threshold_1992}.
\section*{Acknowledgements}
\begin{small}This work has been supported by the \textit{Deutsche Forschungsgemeinschaft} in the framework of the Research
Group 506: "Physics of Nitride-Based Nanostructured Light-Emitting Devices".                                                                            \end{small}
%
%
\providecommand{\othercit}{}
\providecommand{\jr}[1]{#1}
\providecommand{\etal}{~et~al.}

\end{document}